\documentclass[lettersize,journal]{IEEEtran}
\usepackage{cite}
\usepackage{amsmath,amssymb,amsfonts}
\usepackage{algorithm}
\usepackage{algpseudocode}
\usepackage{graphicx}
\usepackage{textcomp}
\usepackage{xcolor}
\usepackage{array}
\usepackage{siunitx}
\usepackage{caption}
\usepackage{subcaption}
\usepackage{booktabs} 
\usepackage{multirow} 
\usepackage{array}
\def\BibTeX{{\rm B\kern-.05em{\sc i\kern-.025em b}\kern-.08em
    T\kern-.1667em\lower.7ex\hbox{E}\kern-.125emX}}
\usepackage{balance}
\usepackage{xurl}
\usepackage[utf8]{inputenc}
\UseRawInputEncoding

\begin{document}
\title{ALCS: An Adaptive Latency Compensation Scheduler for Multipath TCP in \\
Satellite-Terrestrial Integrated Networks}
\author{Lin Wang, Ze Wang, Zeyi Deng, Jingjing Zhang, \IEEEmembership{Member,~IEEE,} and Yue Gao, \IEEEmembership{Fellow,~IEEE}}

\maketitle

\begin{abstract}

The Satellite-Terrestrial Integrated Network (STIN) enhances end-to-end transmission by simultaneously utilizing terrestrial and satellite networks, offering significant benefits in scenarios like emergency response and cross-continental communication. Low Earth Orbit (LEO) satellite networks offer reduced Round Trip Time (RTT) for long-distance data transmission and serve as a crucial backup during terrestrial network failures. Meanwhile, terrestrial networks are characterized by ample bandwidth resources and generally more stable link conditions.
Therefore, integrating Multipath TCP (MPTCP) into STIN is vital for optimizing resource utilization and ensuring efficient data transfer by exploiting the complementary strengths of both networks.
However, the inherent challenges of STIN, such as heterogeneity, instability, and handovers, pose difficulties for traditional multipath schedulers, which are typically designed for terrestrial networks. We propose a novel multipath data scheduling approach for STIN, Adaptive Latency Compensation Scheduler (ALCS), to address these issues. ALCS refines transmission latency estimates by incorporating RTT, congestion window size, inflight and queuing packets, and satellite trajectory information. It further employs adaptive mechanisms for latency compensation and proactive handover management.
It further employs adaptive mechanisms for latency compensation and proactive handover management. Implemented in the MPTCP Linux Kernel and evaluated in a simulated STIN testbed, ALCS outperforms existing multipath schedulers, delivering faster data transmission and achieving throughput gains of 9.8\% to 44.0\% compared to benchmark algorithms.
\end{abstract}

\begin{IEEEkeywords}
Multipath TCP, Satellite-Terrestrial Integrated Networks, LEO, RTT, Handover
\end{IEEEkeywords}

\maketitle

\section{Introduction}\label{intro}


In recent years, the Satellite-Terrestrial Integrated Network (STIN) has emerged as a promising solution to meet the growing demand for seamless and ubiquitous connectivity\cite{yao2018space,8946626}. By leveraging the complementary strengths of satellite and terrestrial communication systems, STIN offers a robust and comprehensive approach to achieving global coverage while ensuring uninterrupted, high-quality communication. Low Earth Orbit (LEO) satellite networks, in particular, provide extensive global coverage, enabling connectivity in remote or underserved regions such as oceans, deserts, mountains, and polar regions---locations where terrestrial networks are not feasible \cite{8610423,8473481}. At the same time, terrestrial networks contribute higher bandwidth and more stable link conditions, particularly in densely populated areas where high-capacity, low-latency connections are crucial. This synergy enhances the overall performance and reliability of STIN.

The integration of both satellite and terrestrial links for data transmission within STIN delivers significant benefits. In cases of terrestrial network disruptions---whether caused by natural disasters or other unforeseen circumstances---satellite networks can serve as a critical fallback, ensuring the continuity of data transmission. Moreover, for long-distance communications, LEO satellites provide a distinct advantage by reducing Round Trip Time (RTT) compared to terrestrial networks. For example, the RTT between Chicago and Tokyo via Inter-Satellite Links (ISLs) is approximately 80 ms, which is 50 ms faster than the current Internet route and 20 ms quicker than the optimal fiber path \cite{handley2019using}.

The simultaneous utilization of satellite and terrestrial networks within STIN can be effectively managed through the deployment of the Multipath TCP (MPTCP) protocol \cite{rfc6182, rfc8684}. MPTCP enables concurrent data transmission over multiple network paths, significantly enhancing both throughput and transmission reliability. This approach has been widely adopted in various terrestrial networks, as demonstrated in \cite{dems, daps, blest, ecf, stms, sttf}. 
At the same time, MPTCP has also shown increasing potential in satellite communication environments \cite{SDN-enabled-LEO, du2015multipath}, optimizing resource usage and ensuring resilient data transmission across diverse network paths.

However, research on MPTCP data scheduling specifically for satellite networks remains limited. The deployment of MPTCP within STIN introduces several challenges due to the network's inherent complexities. These include significant differences in path characteristics such as latency, bandwidth, and packet loss rates, high temporal variability in network conditions, and the constant mobility of LEO satellites. These factors make it difficult to maintain stable and efficient data transmission, posing three primary challenges for MPTCP implementation within STIN:
\begin{itemize}
	\item \textbf{Severe heterogeneity}. The path characteristics can vary significantly between satellite and terrestrial paths in the STIN environment. In addition to significant RTT discrepancies, satellite links often have lower bandwidth compared to terrestrial paths \cite{1677596}. This pronounced path heterogeneity can lead to out-of-order packet delivery and receiver buffer blocking, degrading overall throughput and transmission efficiency.
	\item \textbf{High volatility}. LEO satellite networks, such as Starlink, are highly dynamic, with significant temporal variability in path characteristics. Taking Starlink as an example, the latency variation reaches around 3.8 times that of terrestrial networks while the throughput standard deviation achieves 50.71\% \cite{2023-starlink-measurement}. Moreover, Ground-to-Satellite Links (GSL) are sensitive to weather conditions and signal interference,  further complicating network stability and predictability.
    \item \textbf{Frequent Handover}. Due to the constant movement of LEO satellites, frequent handovers are inherent in satellite networks \cite{mptcp-handover}. These handovers can result in packet drops and increased retransmissions, negatively impacting transmission efficiency and overall performance.
\end{itemize}

The distinctive characteristics of STIN present significant challenges for traditional MPTCP scheduling algorithms.
Due to the limited research on MPTCP data scheduling specifically for satellite networks, the schedulers listed here are primarily designed for terrestrial networks---where paths typically consist of cellular or Wi-Fi connections---including the default minRTT scheduler\cite{mptcp-linux-kernel}, RoundRobin (RR) scheduler\cite{mptcp-linux-kernel}, Delay-Aware Packet Scheduler (DAPS)\cite{daps}, BLocking ESTimation-based scheduler (BLEST)\cite{blest}, and Earliest Completion First (ECF) scheduler\cite{ecf}. 
The minRTT scheduler prioritizes subflows with the lowest RTT, often underestimating the inherent delays of satellite links. Similarly, the RR scheduler distributes traffic equally across paths, failing to consider the substantial differences in subflow capacities. BLEST, which adjusts traffic based on observed network conditions, may struggle to react swiftly to the rapid fluctuations in STIN. Even the more sophisticated ECF scheduler, which considers factors beyond RTT to estimate completion times, can falter under STIN’s frequent handovers and dynamic satellite link conditions. These limitations highlight the need for specialized scheduling strategies that address the unique challenges of STIN to optimize MPTCP performance in such integrated and complex network environments.

In this paper, we propose an Adaptive Latency Compensation Scheduler (ALCS), a novel multipath packet scheduler specifically designed for STIN. ALCS aims to minimize data transmission time, increase the overall network throughput, and enhance transmission reliability across satellite and terrestrial network paths. 
The core innovation of ALCS lies in its ability to adaptively compensate for estimated packet transmission delays in STIN while proactively deactivating the satellite path prior to handovers. This approach takes into account RTT fluctuations and satellite movement, delivering more reliable and efficient data transfer. By optimizing the use of satellite network resources alongside terrestrial networks, ALCS allows for smoother transitions and enhanced performance in dynamic STIN environments.

The main contributions of this paper are summarized as follows:
\begin{itemize}
    \item We introduce an MPTCP data transmission framework tailored for STIN and analyze the performance degradation of traditional MPTCP schedulers when deployed in this environment.    
    \item We design a novel multipath scheduler, ALCS, that addresses path heterogeneity, RTT fluctuations, and handovers caused by satellite movement. By evaluating the transmission delays, ALCS selects the most appropriate path for the current data segments, ensuring reliable and efficient delivery in rapidly changing network conditions.
    \item We implement the ALCS scheduler in the Linux Kernel. Additionally, we build a flexible and scalable testbed using Docker containers and Python scripts, enabling a thorough study of different scheduler behaviors under diverse STIN conditions.
    \item We evaluate the performance of ALCS against several MPTCP schedulers, including minRTT, RR, BLEST, and ECF, in a custom-built experimental testbed. The results demonstrate that our proposed ALCS improves overall performance ranging from 9.8\% to 44.0\% across various metrics.
\end{itemize}

The remainder of this paper is structured as follows. Section \ref{bg} provides the necessary background on the MPTCP protocol in the context of STIN, highlights the unique characteristics of STIN, and examines the performance limitations of traditional schedulers in the STIN environment. In Section \ref{design}, we present the design of our proposed ALCS scheduler. Implementation details and the experimental testbed setup are provided in Section \ref{testbed}, followed by evaluations of ALCS and other scheduling algorithms in Section \ref{results}. Discussions, limitations, and future work are given in Section \ref{discussion}. The relevant related work is reviewed in Section \ref{related-work}, and conclusions are drawn in Section \ref{conclusion}.



\section{Implementation and Challenges of MPTCP in STIN} \label{bg}

In this section, we first present the MPTCP communication diagram for STIN, followed by an overview of one main inherent characteristic of STIN that challenges the effectiveness of MPTCP. Lastly, we analyze the performance degradation seen in various conventional MPTCP scheduling algorithms.

\begin{figure}[t!]
    \centering
    \includegraphics[width=1\linewidth]{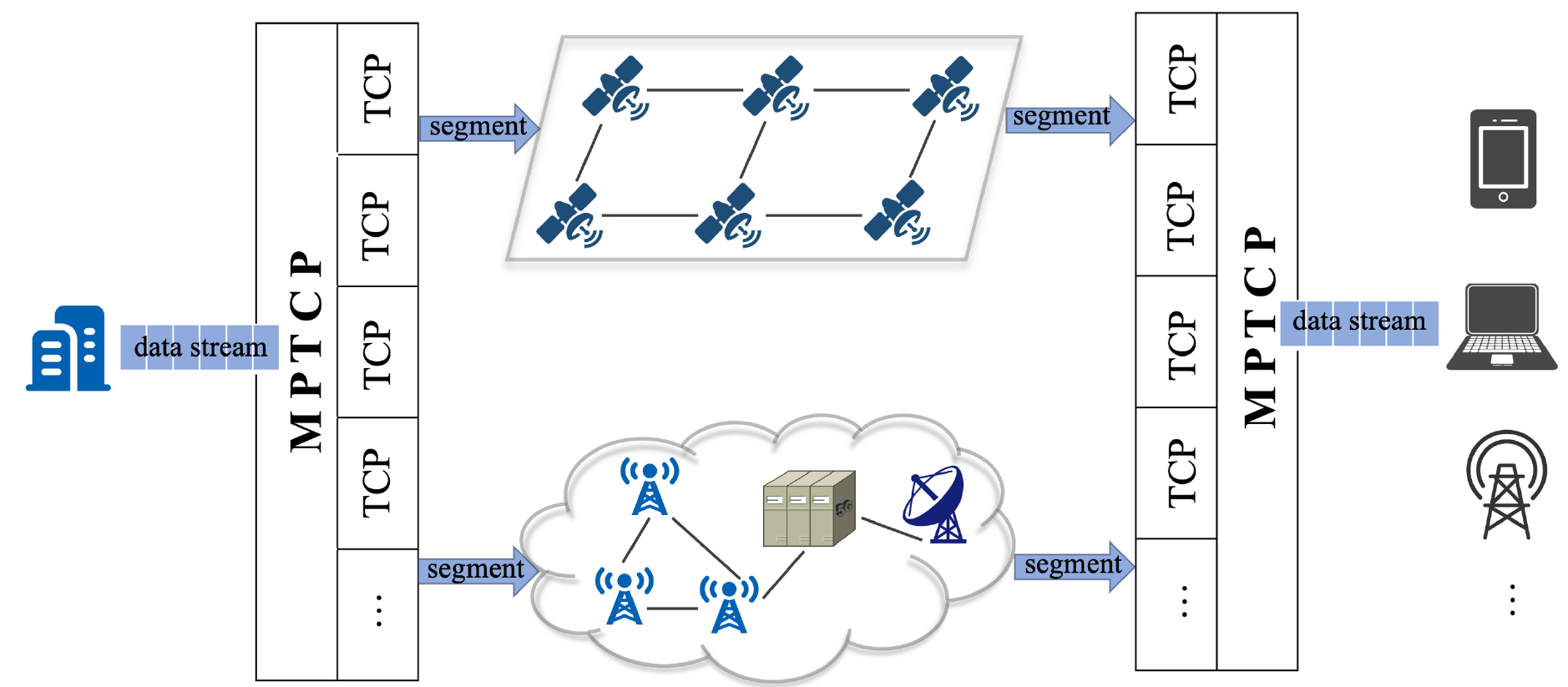}
    \caption{MPTCP data transmission diagram for STIN.}
    \label{mptcp-structure}
\end{figure}

\subsection{MPTCP for STIN}


The MPTCP protocol operates between the transport and application layers, splitting packets into multiple subflows for transmission across different available paths. Each subflow functions independently, resembling a standalone TCP flow. The MPTCP communication process primarily involves connection initialization, the establishment of available subflows, data transmission, and connection termination.

Fig.~\ref{mptcp-structure} illustrates the MPTCP architecture within a STIN, leveraging both satellite and terrestrial networks concurrently. The key communication steps within this STIN context are detailed as follows:

\begin{itemize}
    \item[1] MPTCP first establishes an initial TCP connection, similar to the traditional single TCP connection setup process, which includes the standard three-way handshake. This step is essential for configuring communication parameters and capabilities between the source and destination.
    \item[2] Upon completing the initial connection, MPTCP identifies all available sub-paths, which can be established across different network interfaces, physical routes, and devices. In the case of multiple end-to-end paths, such as terrestrial and satellite links in STIN, corresponding subflows are created. These subflows are used for parallel data transmission, improving reliability and efficiency.
    \item[3] 
    Following the establishment of subflows, MPTCP transmits data concurrently across the different paths. A data scheduler is responsible for selecting the optimal subflow for packet transmission. Consequently, designing an efficient and robust scheduler is critical for ensuring effective data transmission within STIN.
    \item[4] After data transmission is completed, MPTCP terminates the connection, following a procedure similar to the traditional TCP termination process.
\end{itemize}


As emphasized in step 3, efficient multipath data scheduling is critical to improving the performance and effectiveness of MPTCP in real-world networks. By intelligently distributing data traffic based on path characteristics, congestion levels, and application requirements, multipath scheduling algorithms can optimize resource utilization and significantly enhance the end-user experience. However, several inherent attributes of STIN pose significant challenges to traditional data scheduling methods primarily designed for terrestrial networks.

\begin{figure}[t!]
    \centering
    \includegraphics[width=0.9\columnwidth]{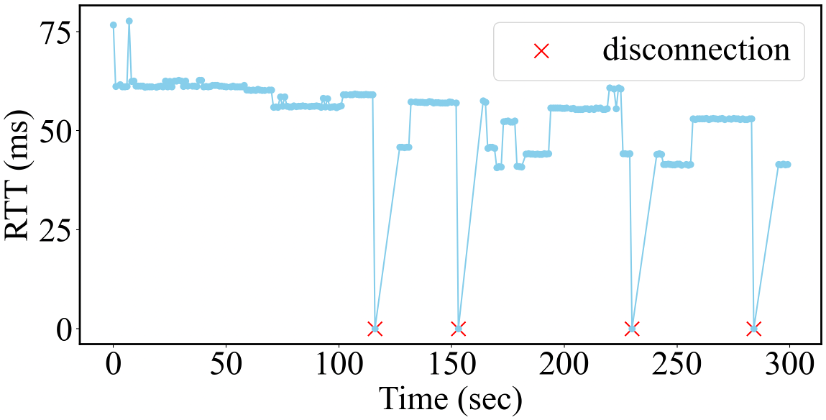}
    \caption{Latency dynamics in a simulated STIN environment.}
    \label{fig:sat-rtt}
\end{figure}

\subsection{Characteristics of STIN}


As discussed in Section \ref{intro}, a major challenge in STIN is the handover issue caused by satellite movement. Frequent handovers between satellites and ground stations, or across satellites, can have a substantial impact on link conditions. For example, in cases where a communication session depends on a single end-to-end satellite path, a handover may result in disconnection, disrupting the continuity of transmission. Conversely, in systems with multiple satellite paths, ISL handovers can introduce fluctuations in RTT and bandwidth variability, which may degrade the overall quality of service.


\begin{figure}[t!]
    \centering
    \includegraphics[width=0.65\columnwidth]{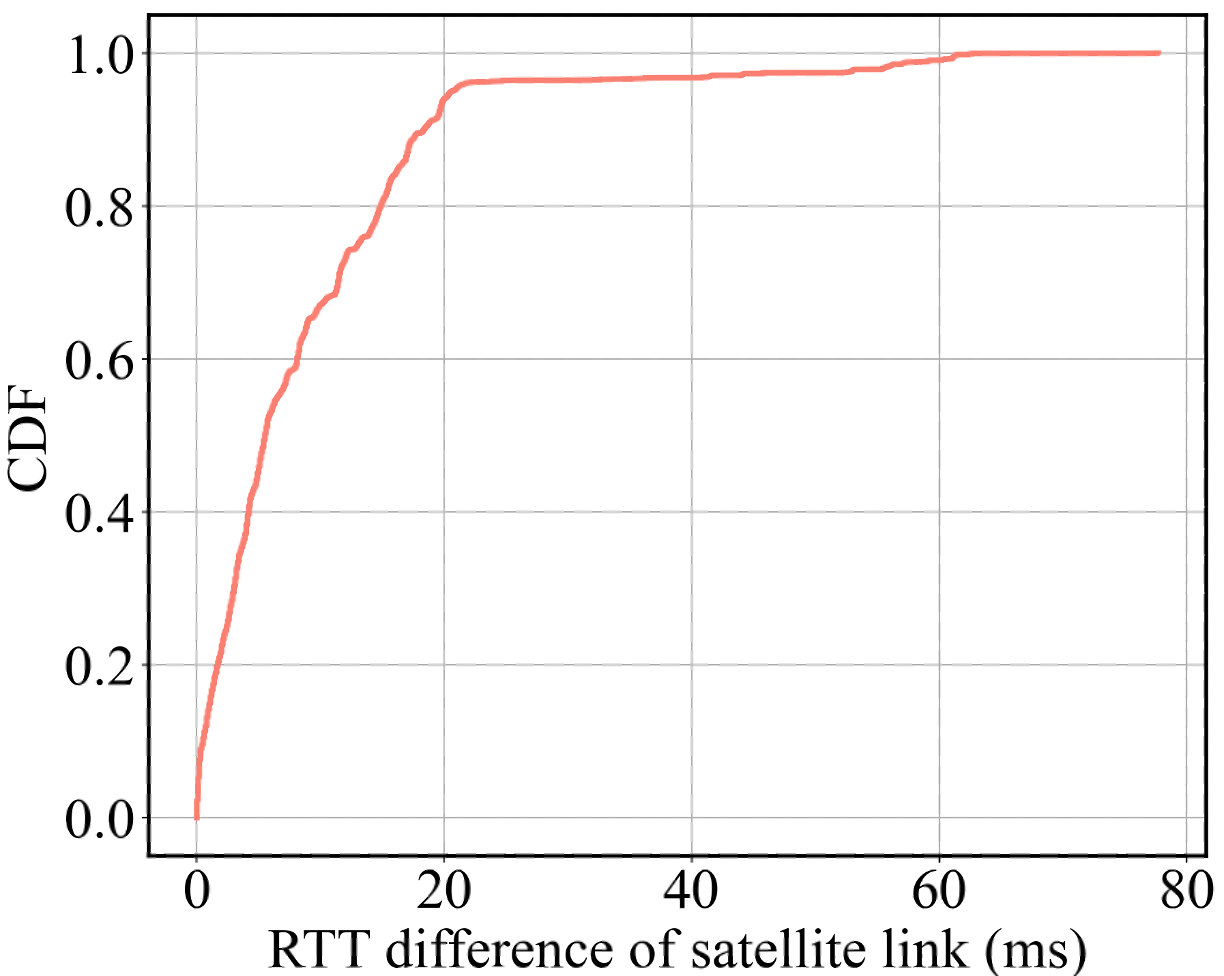}
    \caption{CDF of the difference between the maximum and minimum RTTs of the satellite path.}
    \label{fig:sat-rtt-cdf}
\end{figure}


Fig. \ref{fig:sat-rtt} presents the RTT fluctuations of a satellite connection observed in our custom-built testbed (see Section \ref{testbed} for details). To effectively capture the RTT characteristics within a limited test period, we accelerate the satellite movement in the testbed. This approach enables a 300-second ping test over long-distance communication, offering insights into RTT performance under conditions that simulate real LEO satellite dynamics at an accelerated rate. As shown in Fig. \ref{fig:sat-rtt}, the predominantly stable and low RTT, averaging around 60 ms, highlights the satellite network's potential for reduced latency in long-distance communications, in contrast to terrestrial networks, which typically exhibit RTT values exceeding 100 ms.


Nevertheless, the RTT of the satellite path exhibits significant fluctuations, including notable spikes labeled as ``disconnection" events. These spikes are primarily attributed to handovers between LEO satellites, particularly when a single satellite path is in use. Such handovers can cause temporary disconnections and RTT values spiking well above the baseline. This recurring pattern of handover-induced spikes undermines the overall quality and reliability of the connection, leading to unpredictable service interruptions and degraded performance. Additionally, there are instances where RTT shifts from approximately 60 ms to 40 ms, which are associated with ISL handovers when multiple satellite paths are active in the test environment.

To illustrate the variability in RTT experienced by the satellite link, we present the Cumulative Distribution Function (CDF) of the difference between the maximum and minimum RTTs observed on the satellite path during the 300-second ping test, as shown in  Fig. \ref{fig:sat-rtt-cdf}. The CDF indicates that RTT variations exceeding 10 ms occur with a probability of approximately 0.4, highlighting the significant likelihood of such fluctuations. Furthermore, there is a small but notable chance of encountering RTT variations exceeding 20 ms, emphasizing the inherent volatility of satellite communication links.

Therefore, addressing the varying RTT challenges through advanced mobility management and multipath transmission schemes is essential to fully leverage the benefits of STIN architectures. Effective solutions can ensure seamless, high-quality connectivity despite the dynamic and constantly evolving network topology.

\subsection{Performance Degradation}

Most existing multipath schedulers rely on assumptions of relatively stable path characteristics and availability. However, frequent satellite handovers in STIN scenarios can drastically alter network conditions, making these assumptions invalid. As a result, traditional scheduling algorithms, which optimize based on time-invariant path metrics, may perform poorly, leading to suboptimal resource utilization and potential service disruptions during communications.

\subsubsection{\textbf{Severe Retransmission due to Handover-induced Disconnection}} We conduct a comprehensive evaluation of various MPTCP schedulers in our custom-built testbed, focusing on a single satellite path scenario. When data packets are transmitted over a satellite path nearing handover, the likelihood of packet loss increases dramatically due to the imminent disconnection. This packet loss triggers a cascade of retransmissions, consuming additional bandwidth and system resources, while also introducing latency. The cumulative effect of these retransmissions can significantly degrade the overall transmission performance, leading to slower data delivery rates and higher error rates.

\begin{table}[t!]
    \centering
    \caption{Packet Retransmission due to LEO Satellite Handovers.}
    \begin{tabular}{m{2.5cm}<{\centering}|m{5.3cm}<{\centering}}
        \hline
        \textbf{MPTCP scheduler} & \textbf{retransmission packets caused by handover / total retransmissions} \\
        \hline
        minRTT & 215 / 230 (93.48\%) \\
        RR               & 400 / 437 (91.53\%) \\
        BLEST            & 810 / 822 (98.54\%) \\
        ECF              & 286 / 296 (96.62\%) \\
        \hline 
    \end{tabular}
    \label{tab:retrans}
\end{table}

As illustrated in Table \ref{tab:retrans}, traditional MPTCP schedulers, including minRTT, RR, BLEST, and ECF, suffer from substantial performance degradation due to packet loss and retransmission triggered by satellite handovers. Notably, a vast majority—between 91\% and 98\%—of all retransmissions stem from these handovers, highlighting the inability of existing MPTCP scheduling strategies to effectively mitigate data loss during satellite link transitions. 

In scenarios where data segments are allocated to a satellite path nearing handover, most of these packets may require retransmission as they are lost during the temporary disconnection of the satellite link. While retransmission over the terrestrial path can help ensure seamless transmission at the application layer, this approach comes at the cost of consuming additional terrestrial bandwidth resources. On the other hand, retransmitting via the satellite path forces the receiver to wait until the link is restored, which can lead to significant delays and out-of-order delivery.
This trade-off underscores the importance of proactively preventing packet loss due to handovers, rather than relying on reactive retransmission strategies.


\subsubsection{\textbf{Performance Degradation Still Occurs in Non-Disconnection Scenarios}}

To evaluate the impact of STIN's inherent heterogeneity and satellite link dynamics, we also analyze the performance of various MPTCP schedulers in scenarios where satellite link disconnections are absent throughout the data transmission process.

Table \ref{tab:performance-degradation} presents a comparative analysis of four MPTCP schedulers under these conditions.
We assess the transmission completion time and average throughput while transferring a 100MB file using these schedulers, ensuring no handover events occur during the process. It is observed that the default minRTT scheduler has the longest transfer duration and the lowest overall throughput among the four schedulers. Meanwhile, the RR, BLEST, and ECF schedulers exhibit similar performance, with transmission completion times of approximately 48 seconds and average throughputs of around 2.319MB/s. This suggests that ECF and BLEST may not maintain their advantage in the highly dynamic and heterogeneous STIN environment. The reliance of ECF and BLEST on RTT estimates, which are prone to significant fluctuation in satellite paths, likely impacts the accuracy of their estimations and, consequently, their effectiveness.

Thus, it is crucial to develop multipath data schedulers specifically designed to address the unique challenges of STIN environments.

\begin{table}[t!]
    \centering
    \caption{Performance Comparison in Non-handover STIN Scenarios.}
    \begin{tabular}{m{2.5cm}<{\centering}|m{2.1cm}<{\centering}|m{3.1cm}<{\centering}}
        \hline
        \textbf{MPTCP scheduler} & \textbf{completion time (s)} & \textbf{average throughput (MB/s)} \\
        \hline
        minRTT & 56.779 & \num{1.988} \\
        RR               & 48.462 & \num{2.319} \\
        BLEST            & 48.453 & \num{2.319} \\
        ECF              & 48.018 & \num{2.319} \\
        \hline 
    \end{tabular}
    \label{tab:performance-degradation}
\end{table}

\section{Algorithm Design}\label{design}

This section details the design of our proposed scheduling algorithm, ALCS. We begin with an overview of ALCS, outlining its foundational principles and objectives. Following this, we examine the three pivotal components that form the core of the ALCS framework.


\subsection{Design Overview}

To elaborate, we first define the following variables. The variable $S_f$ denotes the fastest subflow, while $S_s$ represents the slower subflow among the available paths. $rtt_f$ and $cwnd_f$ refer to the RTT and CWND of $S_f$, whereas $rtt_s$ and $cwnd_s$ are for $S_s$. $\Delta_f$ and $\Delta_s$ denote the variations of $rtt_f$ and $rtt_s$, respectively. Furthermore, $tl_f$ and $tl_s$ denote the estimated transmission latencies for $S_f$ and $S_s$, including any waiting time and the transmission delay for data segments scheduled over these subflows. $\sigma_f$ and $\sigma_s$ are latency compensation factors used to refine the estimated latencies. The variable $T_u$ denotes the update interval for revising the compensation factors $\sigma_f$ and $\sigma_s$. 

\begin{figure}[t!]
    \centering
    \includegraphics[width=1\columnwidth]{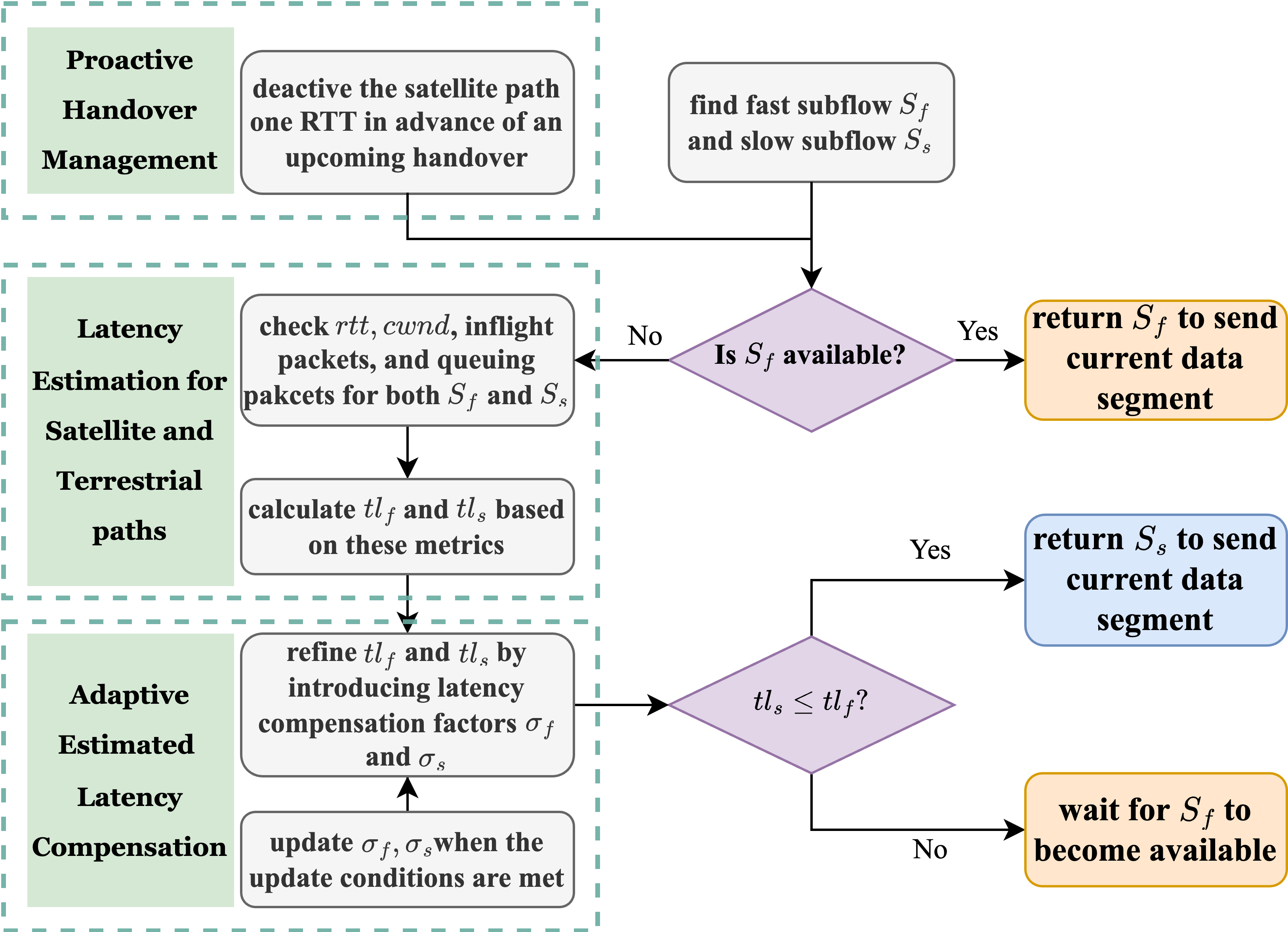}
    \caption{Flowchart diagram of the proposed ALCS.}
    \label{fig:ALCS-flowchart}
\end{figure}



The general workflow of the ALCS scheduling algorithm is outlined in Fig. \ref{fig:ALCS-flowchart} and can be summarized as follows: 1) Identify all available subflows, including both satellite and terrestrial paths within the STIN environment, designating the faster path as $S_f$ and the slower one as $S_s$; 2) Return $S_f$ if it has available CWND capacity; 3) Otherwise estimate the transmission latencies $tl_f$ and $tl_s$ for both $S_f$ and $S_s$, using path metrics $rtt_f, cwnd_f, rtt_s, cwnd_s$, along with inflight packet counts and queuing lengths; 4) Refine the estimated latencies by incorporating latency compensation factors $\sigma_f$ and $\sigma_s$, which are updated at each update interval $T_u$; 5) Based on these calculations, determine whether to proceed with $S_f$ immediately or defer transmission until $S_f$ becomes available again.
It is important to note that in the long-distance communication scenarios under study, the satellite path utilizing ISLs typically exhibits a lower RTT than the terrestrial path. Consequently, $S_f$ predominantly signifies the satellite path in this context.

The ALCS algorithm is designed to capitalize on the complementary strengths of satellite and terrestrial networks, aiming to enhance the reliability and efficiency of end-to-end communications. To this end, ALCS builds on the strengths of existing scheduling methods while introducing enhancements tailored to the specific challenges of the STIN environment. These challenges, as outlined in Section \ref{intro}, include significant network heterogeneity, high volatility in link quality, and frequent handovers. As depicted in Figure \ref{fig:ALCS-flowchart}, ALCS is structured around three key components: latency estimation for satellite and terrestrial paths, adaptive estimated latency compensation to fine-tune performance, and proactive handover management to ensure uninterrupted connectivity.
The strategies employed by each of these three components to address the corresponding challenges are listed below.
\begin{itemize}
    \item \textbf{Latency Estimation for Satellite and Terrestrial Paths}: Given the inherent differences between LEO satellite networks and terrestrial networks, ALCS implements a mechanism to estimate the transmission latency of data segments across both paths. This estimation factors in various network dynamics, such as RTT, CWND, inflight packets, and queue lengths, enabling the scheduler to make informed decisions based on real-time path conditions.   
    \item \textbf{Adaptive Estimated Latency Compensation}: To address performance issues arising from fluctuating satellite links, ALCS introduces an adaptive latency compensation mechanism. This component updates the latency compensation factors by periodically comparing the actual and estimated latencies. In this manner, it refines the latency estimation, thereby mitigating the adverse effects caused by the variability of link conditions.    
    \item \textbf{Proactive Handover Management}: Leveraging the predictable movement patterns of LEO satellites, ALCS employs proactive handover management. By maintaining an updated schedule of upcoming handover events, the algorithm can preemptively disable satellite links ahead of time, preventing packet loss and retransmissions during handovers. This strategy enhances overall transmission efficiency and reduces unnecessary resource consumption.    
\end{itemize}

Design details of the three components are elaborated in the following parts.


\subsection{Latency Estimation for Satellite and Terrestrial Paths}

We assume that $k$ packet segments in the MPTCP connection level send buffer are waiting to be scheduled to the best available subflow. Similar to the default minRTT scheduler, if the fastest subflow \textemdash characterized by the minimal RTT \textemdash has available CWND space, ALCS would readily assign the packets to that subflow. However, in cases where the faster subflow $S_f$ lacks sufficient CWND space, ALCS would not allocate the data packets in the send buffer to the slower subflow $S_s$ immediately.

Instead, in cases where $cwnd_f$ is insufficient for the pending data, carefully balances the trade-off between latency and throughput before deciding on which subflow to send the packets.
Beyond just RTT and CWND, ALCS also considers the number of inflight packets and the queue length in the send buffer for each subflow. These additional factors allow ALCS to build a more comprehensive view of each subflow's state, leading to more informed decision-making. Once this analysis is completed, ALCS computes the end-to-end latency for transmitting the data segment over both subflows, denoted as $tl_f$ and $tl_s$ for $S_f$ and $S_s$, respectively. 

Given the assumption that there are $k$ data segments awaiting allocation, the calculation of $tl_f$ and $tl_s$ is given as
\begin{align}
    &tl_f=\frac{k+i_f+q_f}{cwnd_f}*(rtt_f+\Delta_f), \label{eq:ori-estimate-tlf} \\
    &tl_s=\frac{k+i_s}{cwnd_s}*(rtt_s+\Delta_s).\label{eq:ori-estimate-tls}
\end{align}
Here, $i_f$ and $i_q$ represent the number of inflight packets for subflows $S_f$ and $S_s$, respectively, and $q_f$ indicates the number of packets queued in the send buffer of $S_f$. The subflow expected to deliver the segment in the shortest total time is selected for transmission.

\subsection{Adaptive Estimated Latency Compensation}

To address potential discrepancies between estimated and actual RTT, our scheduling approach integrates an adaptive latency compensation mechanism to refine latency estimates and mitigate the negative impact of RTT variability. To elaborate, we introduce a latency compensation factor, $\sigma$, which is dynamically adjusted based on the difference between the estimated transmission latency and the actual observed delay during recent transfer intervals. Initially, $\sigma_f$ and $\sigma_s$ are set to $1.0$ in ALCS. 

Within the ALCS framework, this adjustment process involves systematically assessing the variance between observed and predicted end-to-end transmission latencies at regular intervals. This is achieved by comparing the number of data packets successfully transmitted (denoted as $atp$ for ``actually transmitted packets'') with the expected number of dispatched packets (denoted as $etp$ for ``expected transmitted packets'') within a defined update interval $T_u$.
If $atp > etp$, it indicates that the transmission latency has been overestimated, prompting the compensation factor $\sigma$ to be decreased by multiplying it with an attenuation factor, $\alpha$. Conversely, if $atp < etp$, $\sigma$ is increased by dividing it by $\alpha$. This attenuation mechanism tempers the adjustments to avoid overcorrection, ensuring that latency estimates remain precise without introducing new inaccuracies. Consequently, ALCS strikes a balance between improving latency prediction accuracy and maintaining network robustness under fluctuating conditions.

By employing this approach, Eq. (\ref{eq:ori-estimate-tlf}) and (\ref{eq:ori-estimate-tls}) are reformulated as
\begin{align}
    &tl_f=\frac{k+i_f+q_f}{cwnd_f}*(rtt_f+\Delta_f)*\sigma_f,\label{eq:estimate-tlf} \\
    &tl_s=\frac{k+i_s}{cwnd_s}*(rtt_s+\Delta_s)*\sigma_s.\label{eq:estimate-tls}
\end{align}
$\sigma_f$ and $\sigma_s$ are updated at each $T_u$ interval as
\begin{align}
    &\sigma_f \leftarrow \left\{
    \begin{array}{rcl}\label{eq:update-sigma-f}
    \sigma_f*\alpha  &     & {apt_f > etp_f,}\\
    \sigma_f/\alpha  &     & {apt_f < etp_f.}
    \end{array} \right. \\
    &\sigma_s \leftarrow \left\{
    \begin{array}{rcl}\label{eq:update-sigma-s}
    \sigma_s*\alpha  &     & {apt_s > etp_s,}\\
    \sigma_s/\alpha  &     & {apt_s < etp_s.}
    \end{array} \right.
\end{align}

\subsection{Proactive Handover Management}

Last but not least, to mitigate the handover-induced severe retransmission problem described in Section \ref{bg}, ALCS includes a proactive handover management mechanism based on the predictable trajectories of satellite motion.

Specifically, ALCS maintains a 24-hour schedule of satellite movements to predict handover times accurately. It strategically deactivates the satellite link one RTT before an inter-satellite handover is expected. This preemptive deactivation minimizes disruptions to ongoing data transmissions by ensuring that the satellite path is disabled just before the handover occurs. The one RTT buffer period allows sufficient time for any in-flight data packets to be successfully transmitted, avoiding packet loss during the handover process. 

By leveraging this advanced knowledge of handover timing, ALCS ensures continuous end-to-end connectivity, facilitating seamless handovers at the application layer. This proactive approach improves data transmission reliability and significantly enhances the overall user experience.

\begin{algorithm}[t!]
    \label{ALCS}
    \begin{algorithmic}[0]
    \Procedure{\textbf{Main Process of ALCS}}{}
    \State initialize current timestamp $T_0$, scale ratio $\alpha=0.99$, 
    \State $\sigma_f = \sigma_s = 1.0$.
    \If{a handover is scheduled to occur within one RTT}
    \State deactivate the satellite path
    \EndIf
    \State find fastest subflow $S_f$ with smallest RTT
    \If{$S_f$ is available}
    \State return $S_f$
    \Else
    \State find $S_s$ with the second smallest RTT using the 
    \State \hspace{-1.1em} default MPTCP scheduler
    \State initialize the estimated transmission latency $tl_f=$ 
    \State \hspace{-1.1em} $tl_s=0$ using $S_f$ and $S_s$ respectively
    \If{$t-T_0\geq T_{u}$}
    \State \Call{Update Latency Compensation Factor}{}
    \EndIf
    \State obtain packets $k$ need to be transferred
    \State calculate $tl_f$ and $tl_s$ using Eq (\ref{eq:estimate-tlf}) and (\ref{eq:estimate-tls})
    \If{$tl_f < tl_s$}
    \State return NULL (waiting for $S_f$ to be available)
    \Else
    \State return $S_s$ (using $S_s$ for transfer at once)
    \EndIf
    \EndIf
    \EndProcedure
    \State \text{  }
    \Procedure{\textbf{Update Latency Compensation Factor}}{}
    \State calculate $etp_f$ and $etp_s$ over $S_f$ and $S_s$ during $t-T_0$
    \State obtain $atp_f$ and $atp_s$ over $S_f$ and $S_s$ during $t-T_0$
    \State update $\sigma_f$ and $\sigma_s$ according to Eq (\ref{eq:update-sigma-f}) and (\ref{eq:update-sigma-s})
    \State limit $\sigma_{min} \le \sigma \le \sigma_{max}$
    \State update $T_0=t$
    \EndProcedure
    
    \end{algorithmic}
\end{algorithm}

Details of the scheduling procedures are outlined in the accompanying pseudo-code for the proposed ALCS algorithm.

\section{Experimental Setup}\label{testbed}

In this section, we simulate a simplified STIN testbed using Docker \cite{docker} and Python3. Our proposed ALCS approach is implemented within the MPTCP Linux Kernel \cite{mptcp-linux-kernel} by modifying the MPTCP scheduler module.

\subsection{Experimental Settings}

\begin{figure}[t!]
    \centering
    \includegraphics[width=1\columnwidth]{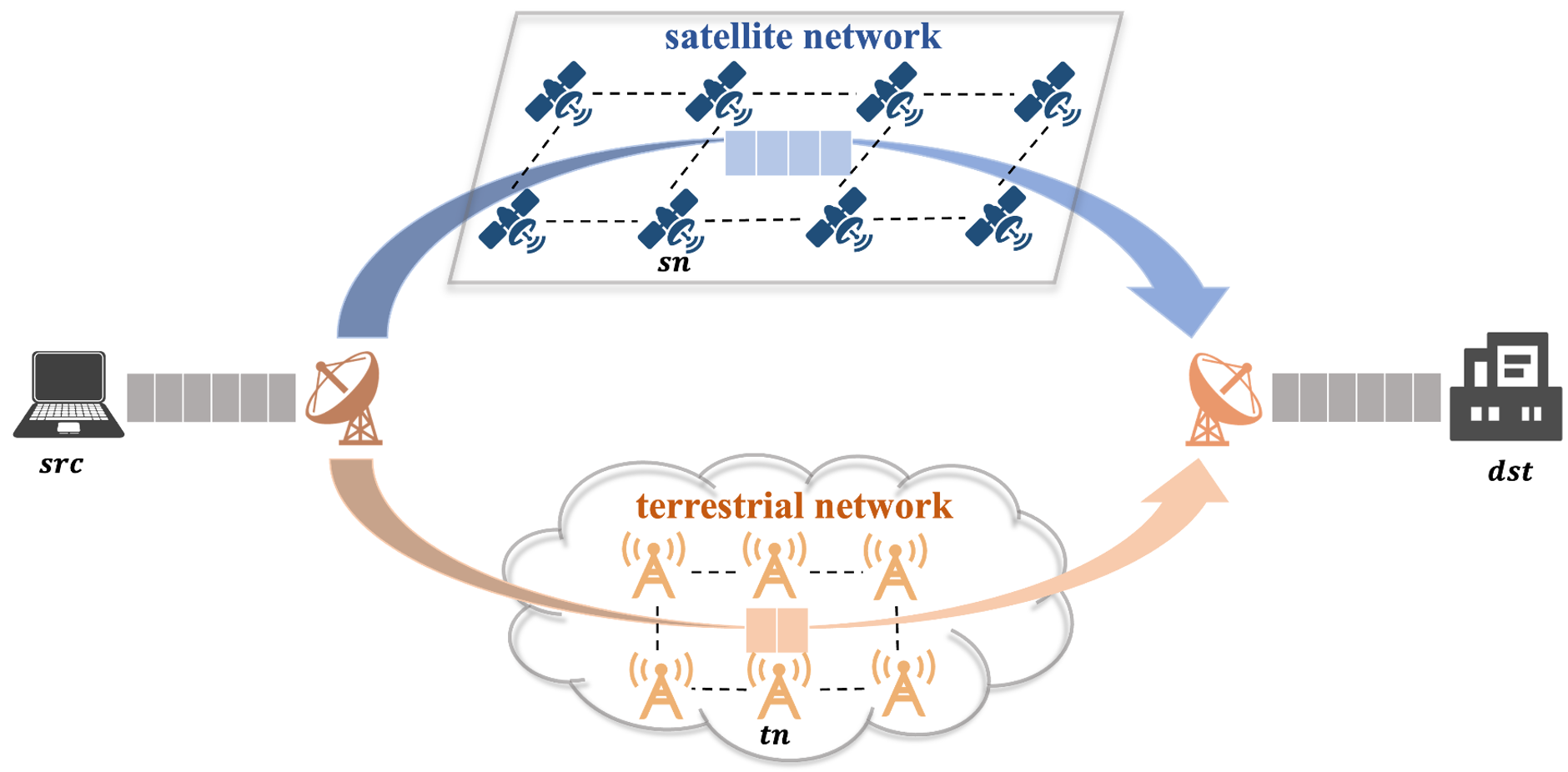}
    \caption{Topology of the simulated STIN.}
    \label{topology}
    \vspace{-10pt}
\end{figure}

The topology of our simplified STIN system is illustrated in Fig. \ref{topology}.
The satellite network comprises multiple interconnected Satellite Nodes ($sn$s), which establish links with both the Source ($src$) and the Destination ($dst$) via ground stations. In the meanwhile, the terrestrial network is depicted by several Terrestrial Nodes ($tn$s), interconnected through ground-based infrastructure, without loss of generality. By utilizing multiple interfaces of both $sn$s and $tn$s, communication between $src$ and $dst$ can effectively employ both satellite and terrestrial networks simultaneously, optimizing data transmission across both paths.

The experimental setup is designed to emulate the operational characteristics of the STIN system. In our configuration, the terrestrial link is specified with a bandwidth of 100 Mbps, and an RTT around 108ms, calculated based on geographical distance. The satellite link, meanwhile, is set with a bandwidth of 50 Mbps, and its RTT varies between 52ms and 68ms due to satellite movement, reflecting the dynamic nature of satellite communication networks. Furthermore, the terrestrial network is assigned a loss rate of 0.01\%. In parallel, to mimic the inherent variability of satellite links, the satellite path is tested under two distinct conditions, with loss rates of 0.01\% and 0.5\%, respectively.

\subsection{Experimental Testbed}

We establish a Docker-based testbed to provide a highly flexible and configurable platform for evaluating the performance of multipath transmission schedulers in the context of STIN. By leveraging containerization, each network node is defined with its own set of network interfaces and interconnections, simulating the diverse paths and varying performance characteristics inherent to satellite and terrestrial network segments. This modular architecture allows for seamless reconfiguration and scaling, facilitating the exploration of a wide range of STIN scenarios.

To create a realistic, controlled, and compatible testbed that accurately simulates dynamic STIN conditions and evaluates the performance of various multipath schedulers, we employ the following core techniques:
\begin{itemize}
    \item \textbf{Two-Line Element (TLE) data for satellite orbits}. The simulation incorporates TLE data from Starlink satellites to ensure an accurate representation of satellite positions and movements. TLE data provides precise orbital information, allowing the testbed to realistically simulate the mobility of $sn$s.    
    \item \textbf{Geospatial coordinates for terrestrial network nodes}. The testbed uses the latitude and longitude coordinates of real-world cities to set the positions of $tn$s, enabling more accurate simulations of communication between ground stations and satellites, and terrestrial nodes. 
    \item \textbf{Accurate delay calculation for satellite and terrestrial links}. Delays for both satellite and terrestrial links are calculated using actual distances. Satellite link delays are derived from TLE data to measure distances between $sn$s and ground stations, while terrestrial delays are based on the geographical distance between $tn$s. This ensures precise latency modeling within the network simulation.
    \item \textbf{Traffic control for network parameter management}. Traffic control for network parameter management. The Linux Traffic Control (TC) utility is used to manage network parameters, such as latency, bandwidth, and packet loss, for both satellite and terrestrial links. TC allows precise control over link characteristics, ensuring realistic simulation of network conditions.  
    \item \textbf{Script-driven simulation of satellite movement and dynamic network conditions}. A script-driven approach is employed to simulate satellite handovers by dynamically establishing and disconnecting network bridges between Docker containers. This technique accurately mirrors satellite trajectories and handover events as satellites move in and out of coverage areas, enabling a realistic assessment of how handovers affect network topology and path availability. Additionally, a Python script recalculates delays for both satellite and terrestrial links in real time, with updated values applied using the TC utility. This ensures that network conditions remain responsive to the continuous changes induced by satellite movement.    
    \item \textbf{MPTCP Linux Kernel v0.96 for multipath communication}. The MPTCP Linux Kernel~\cite{mptcp-linux} version 0.96 underpins the evaluation of multipath schedulers. This version supports the core functionality and APIs required to enable multipath communication. Its compatibility with Linux Kernel Long-Term Support (LTS) release v5.4.230 ensures a stable and widely adopted foundation for the testbed. This environment allows for the reliable and robust evaluation of multipath data transmission.
\end{itemize}

By applying these methodologies, the testbed offers a controlled, reproducible platform for systematically comparing scheduling algorithms across diverse STIN scenarios.

\begin{figure*}[t!]
    \centering
    \vspace{-10pt}

    \subfloat[40 MB, Loss rate: 0.01\%]{
        \includegraphics[width=0.3\textwidth]{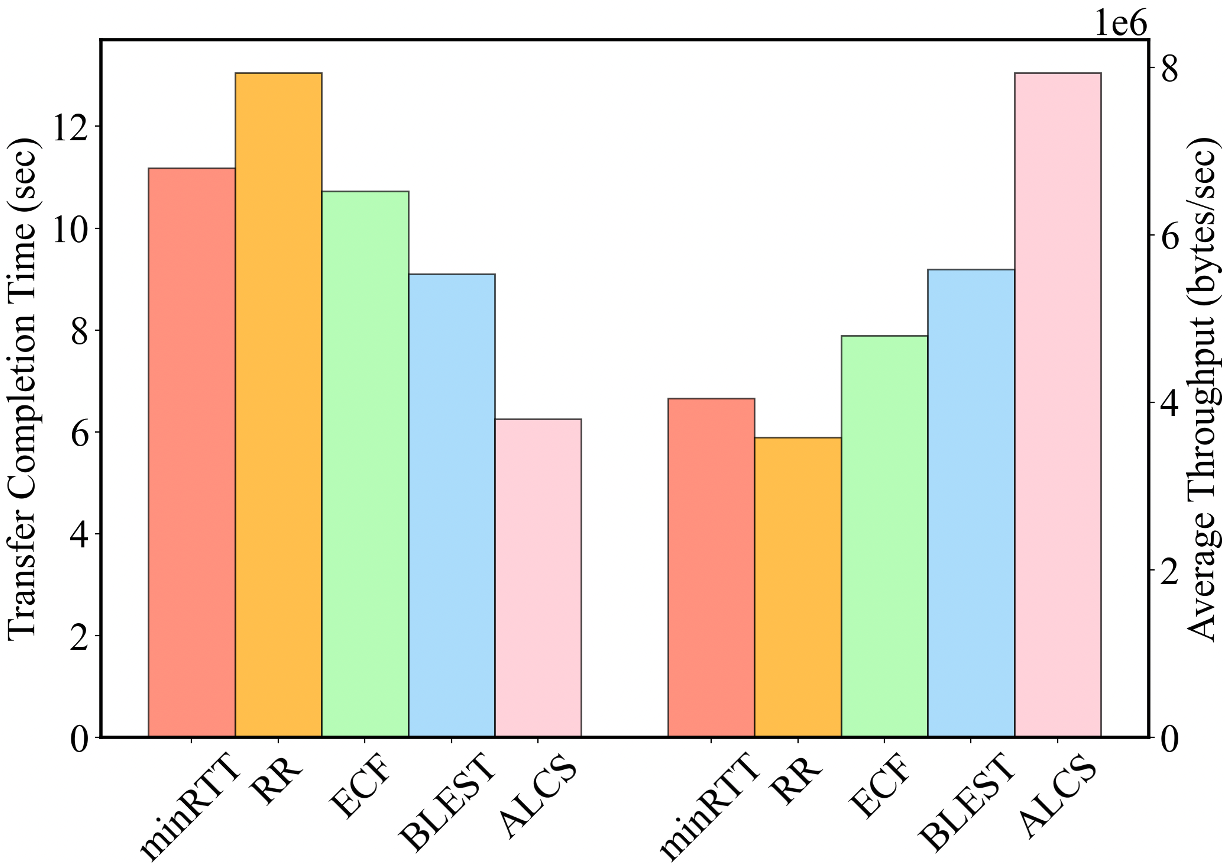}
        \label{fig:throughput-sat0.01-50-40M}
    }
    \hfill
    \subfloat[100 MB, Loss rate: 0.01\%]{
        \includegraphics[width=0.3\textwidth]{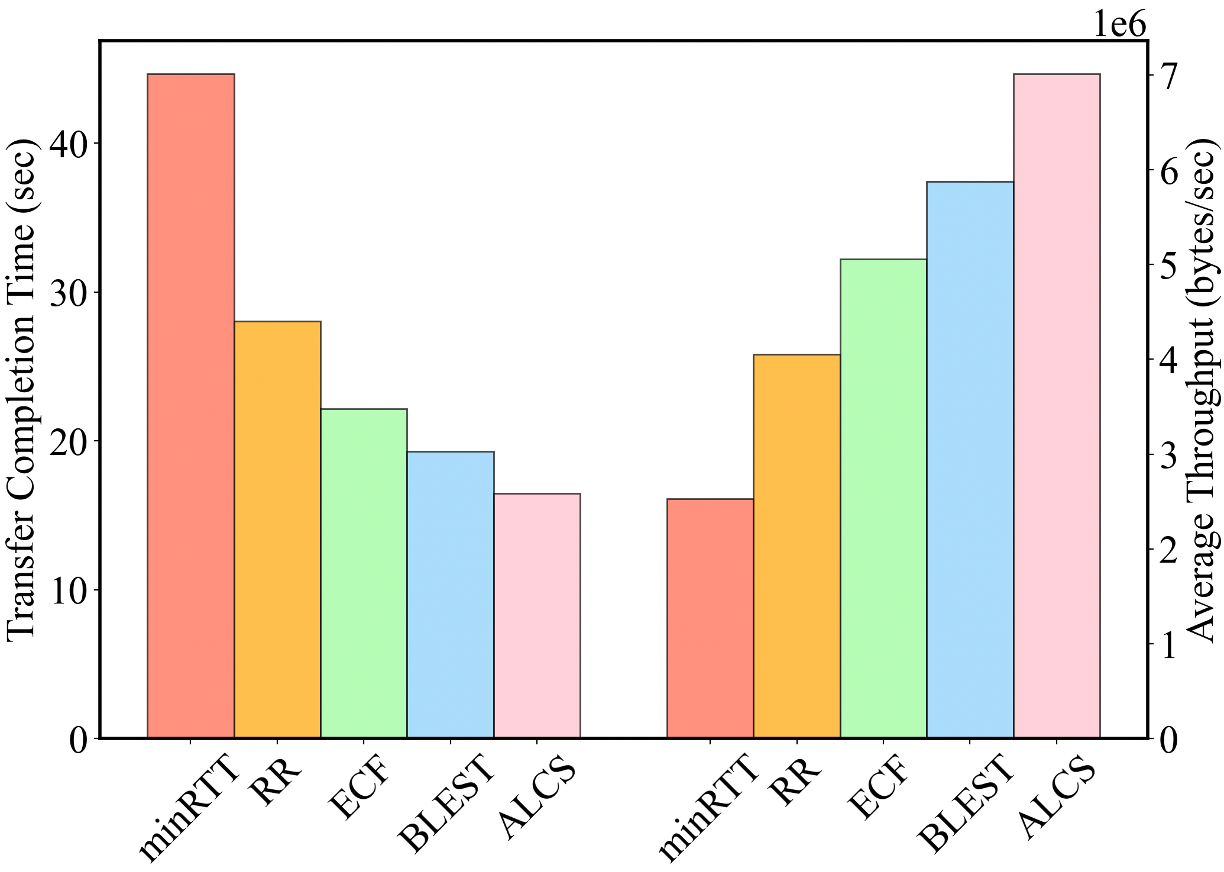}
        \label{fig:throughput-sat0.01-50-100M}
    }
    \hfill
    \subfloat[150 MB, Loss rate: 0.01\%]{
        \includegraphics[width=0.3\textwidth]{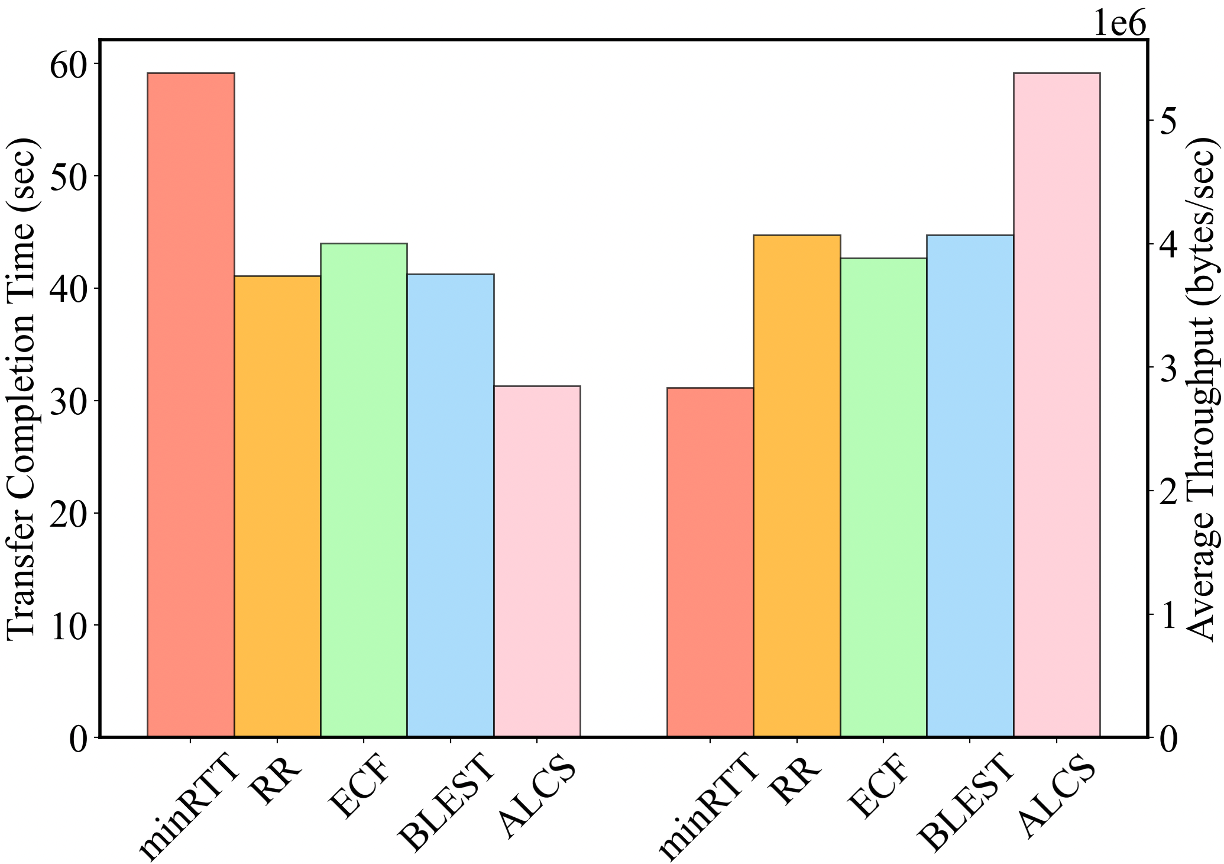}
        \label{fig:throughput-sat0.01-50-150M}
    }

    \vspace{5pt} 

    \subfloat[40 MB, Loss rate: 0.5\%]{
        \includegraphics[width=0.3\textwidth]{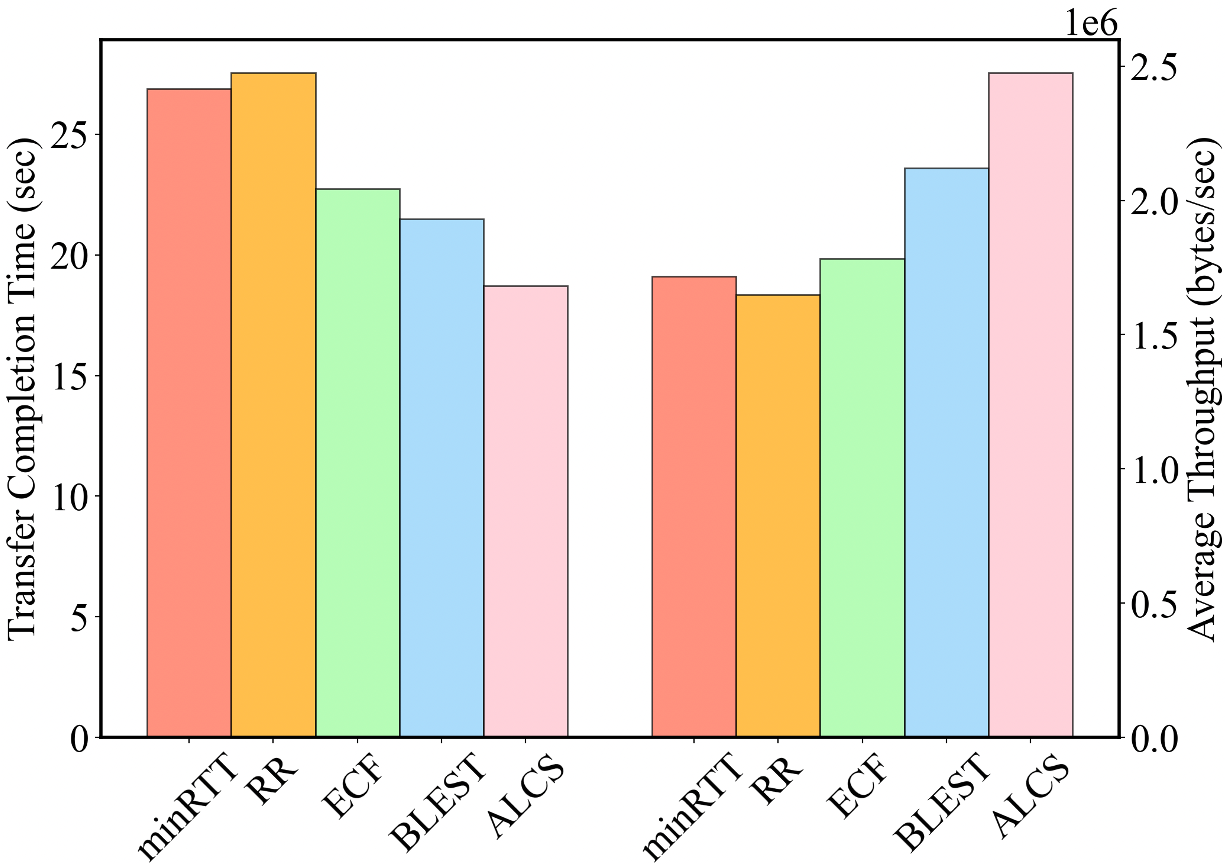}
        \label{fig:throughput-sat0.5-50-40M}
    }
    \hfill
    \subfloat[100 MB, Loss rate: 0.5\%]{
        \includegraphics[width=0.3\textwidth]{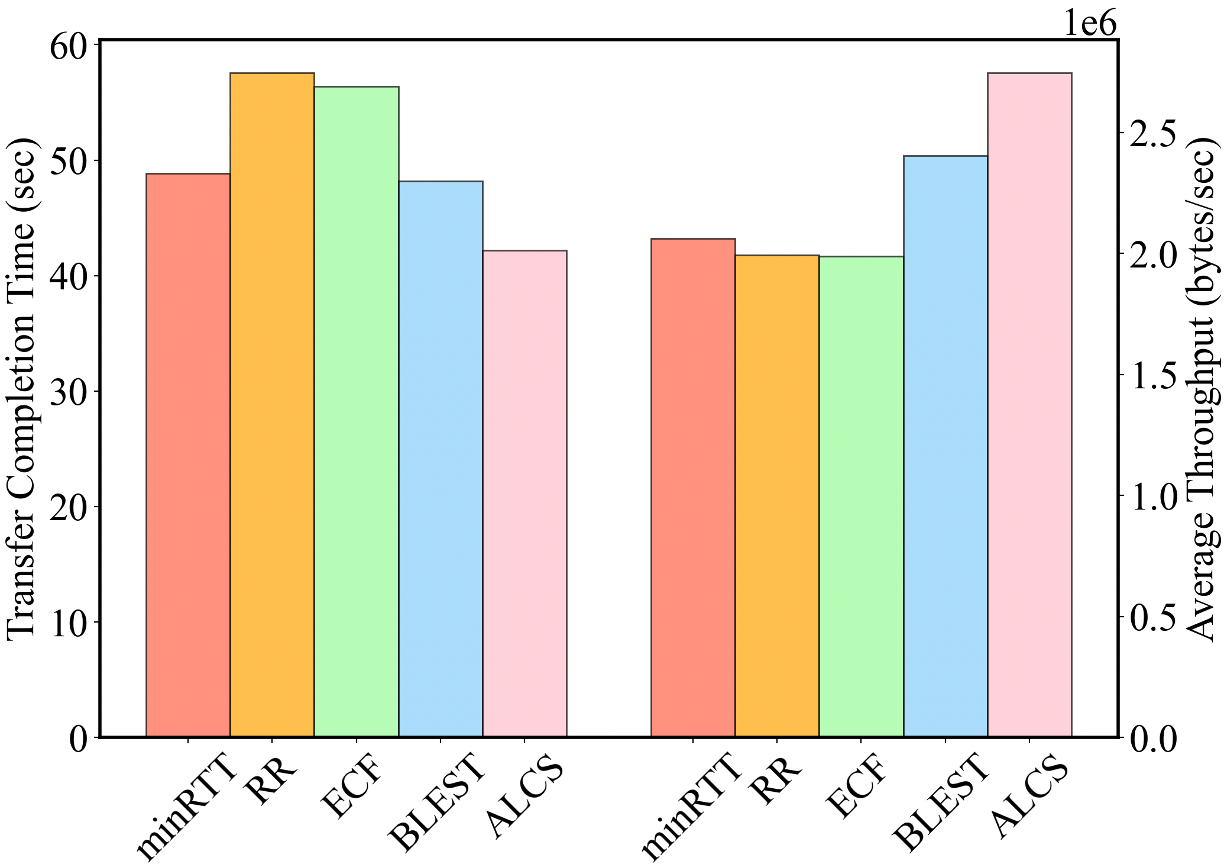}
        \label{fig:throughput-sat0.5-50-100M}
    }
    \hfill
    \subfloat[150 MB, Loss rate: 0.5\%]{
        \includegraphics[width=0.3\textwidth]{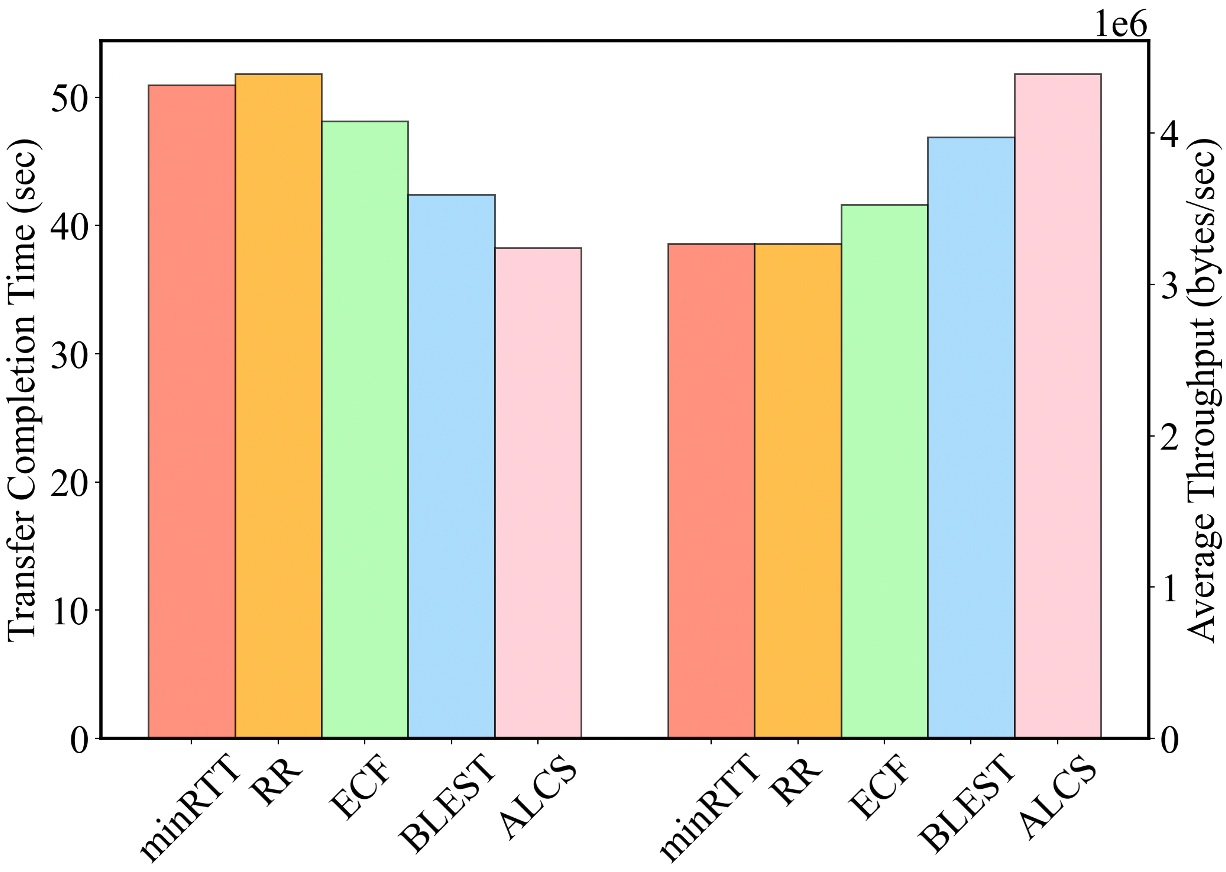}
        \label{fig:throughput-sat0.5-50-150M}
    }

    \caption{
        Comparison of average transfer completion time and throughput for all algorithms 
        when transferring files of 40 MB, 100 MB, and 150 MB at different satellite network loss rates.
    }
    \label{fig:throughput-all}
    
    \vspace{-10pt}
\end{figure*}

\section{Evaluation Results}\label{results}

In this section, we conduct a comprehensive evaluation of the ALCS scheduler within the enhanced simulation environment outlined above. The performance of ALCS is compared against several alternative multipath scheduling approaches, as outlined below:
\begin{itemize}
    \item \textbf{minRTT}~\cite{mptcp-linux-kernel}: This scheduler prioritizes the subflow with the lowest observed RTT for data transmission. While it reduces latency, it may neglect higher-bandwidth paths that could offer superior performance.
    \item \textbf{RR}~\cite{mptcp-linux-kernel}: The RR scheduler distributes traffic evenly across all active subflows, regardless of the current network conditions. Though this ensures a balanced use of resources, it can lead to inefficient network utilization.     
    \item \textbf{ECF}~\cite{ecf}: The ECF scheduler dynamically estimates the completion time for each subflow, prioritizing the subflow expected to complete data transfers fastest by accounting for factors beyond just RTT and CWND.
    \item \textbf{BLEST}~\cite{blest}: The BLEST scheduler estimates the blocking time for each subflow using various network performance metrics. It selects the subflow that promises the highest throughput potential, rather than favoring the path solely based on lower latency.
\end{itemize}

Our evaluation focuses on the following key performance metrics:
\begin{itemize}
    \item \textbf{Transmission Completion Latency}: It measures the time taken for files of a fixed size to travel from $src$ to $dst$ within the simulated STIN. It reflects ALCS's ability to minimize transmission delays, which is crucial for low-latency applications.
    \item \textbf{Overall Throughput}: We evaluate the data transfer rates achieved by each scheduler. This ensures that the latency optimizations provided by ALCS do not compromise throughput, as maintaining high throughput is equally important for efficient data delivery in STIN.
    \item \textbf{Retransmission Rate}: This metric examines the effectiveness of handover management by comparing the retransmission rates of different MPTCP schedulers during handover events, providing insight into their ability to maintain reliable data transmission.
\end{itemize}


Furthermore, we also examine the proportion of data transmitted over satellite paths and analyze the influence of varying the number of subflows on overall performance.

\subsection{Transfer time and network throughput of the schedulers}

We first assess the schedulers by comparing their average throughput and transmission completion time from $src$ to $dst$ within our simulated STIN testbed. These comparisons are visually represented in Fig. \ref{fig:throughput-all}. A more detailed breakdown of ALCS's performance relative to other schedulers across various scenarios is provided in Table \ref{tab:overall-performance}.

Notably, to comprehensively account for the impact of satellite motion, we accelerate the satellite movement during the experiments. Each of the three sets of experiments is conducted at different times, ensuring that the transmission process experiences diverse changes in STIN conditions. However, within each set, two sub-tests---one with a packet loss rate of 0.01\% and the other with 0.5\%---are carried out simultaneously, allowing for a direct comparison of scheduler performance under varying packet loss conditions.

\begin{table*}[t!]
  \centering
  \caption{Overall Performance of ALCS compared to minRTT, RR, ECF, and BLEST in different scenarios.}
    \begin{tabular}{c|c|c|c|c|c}
    \toprule
          &       & \multicolumn{2}{c|}{loss rate of satellite network: 0.01\%} & \multicolumn{2}{c}{loss rate of satellite network: 0.5\%} \\
\cmidrule{2-6}          & MPTCP scheduler & \multicolumn{1}{l|}{completion time (s)} & \multicolumn{1}{l|}{average throughput(MB/s)} & \multicolumn{1}{l|}{completion time (s)} & \multicolumn{1}{l}{average throughput(MB/s)} \\
    \midrule
    \multicolumn{1}{r|}{\multirow{5}[2]{*}{file size: 40M}} & minRTT & 11.183 & 4.05 & 26.885 & 1.72 \\
          & RR & 13.045 & 3.58 & 27.533 & 1.65 \\
          & ECF   & 10.722 & 4.80 & 22.741 & 1.78 \\
          & BLEST & 9.103 & 5.59 & 21.482 & 2.12 \\
          & ALCS  & 6.257 & 7.93 & 18.719 & 2.47 \\
    \midrule
    \multicolumn{1}{r|}{\multirow{5}[2]{*}{file size: 100M}} & minRTT & 44.647 & 2.53 & 48.828 & 2.06 \\
           & RR    & 28.048 & 4.05 & 57.537 & 1.99 \\
          & ECF   & 22.169 & 5.06 & 56.401 & 1.99 \\
          & BLEST & 19.293 & 5.88 & 48.162 & 2.40 \\
          & ALCS  & 16.473 & 7.01 & 42.201 & 2.75 \\
    \midrule
    \multicolumn{1}{r|}{\multirow{5}[2]{*}{file size: 150M}} & minRTT & 59.151 & 2.83 & 50.975 & 3.27 \\
          & RR & 41.102 & 4.07 & 51.816 & 3.27 \\
          & ECF   & 43.981 & 3.88 & 48.159 & 3.53 \\
          & BLEST & 41.239 & 4.07 & 42.424 & 3.97 \\
          & ALCS  & 31.302 & 5.38 & 38.286 & 4.39 \\
    \bottomrule
    \end{tabular}%
  \label{tab:overall-performance}%
\end{table*}%

Fig.~\ref{fig:throughput-sat0.01-50-40M} and \ref{fig:throughput-sat0.5-50-40M} present a comparison of the average throughput and transfer completion time for a 40MB file transmission using ALCS and four other MPTCP schedulers, under satellite network loss rates of 0.01\% and 0.5\%. It is evident that the ALCS scheduler consistently achieves shorter completion times and higher throughput across both loss scenarios, outperforming the alternative scheduling approaches. This trend persists in the transmission of larger files, as shown in Fig.~\ref{fig:throughput-sat0.01-50-100M},~\ref{fig:throughput-sat0.5-50-100M} and Fig.~\ref{fig:throughput-sat0.01-50-150M},~\ref{fig:throughput-sat0.5-50-150M}, for 100MB and 150MB files, respectively.

As expected, an increase in the satellite network's loss rate leads to a reduction in average throughput and longer transfer times for all schedulers across the three file sizes. The higher loss rate reflects increased link instability, prompting a heavier reliance on the terrestrial path, which has a higher RTT than the satellite link. This shift contributes to the observed decline in throughput. Despite this, ALCS consistently outperforms its counterparts, maintaining higher throughput and demonstrating superior resilience to packet loss. This highlights ALCS's effectiveness in managing the dynamic network conditions typical of STIN, ensuring both robust and efficient data transmission.


As the file size increases from 40M to 150M, all schedulers exhibit a predictable rise in completion time and a corresponding decline in average throughput. However, the performance gap between ALCS and the other schedulers becomes more pronounced with larger file sizes. The greater the file size, the longer the transfer duration, reflecting the growing complexities and dynamics in the transmission process. Under such circumstances, traditional MPTCP scheduling algorithms tend to suffer from performance degradation. For instance, in the 150MB file transfer scenario, both ECF and BLEST demonstrate performance comparable to RR and minRTT. Notably, in the case of a 0.01\% satellite link loss rate, ECF requires even more time than RR, indicating suboptimal performance in this specific situation.

In contrast, ALCS consistently outperforms the other schedulers, demonstrating a clear advantage across all file sizes and loss rates. This superior performance suggests that ALCS is not only more scalable but also more capable of handling larger data transfers within STIN environments. Its robustness in managing varying data loads highlights its effectiveness in adapting to different network conditions and maintaining efficient data transmission.

To summarize, our proposed ALCS consistently achieves higher average throughput and shorter transfer completion time compared to the other MPTCP schedulers, across all three file sizes and both loss rate conditions. For smaller files, such as 40MB, and with a lower loss rate of 0.01\%, ALCS delivers a significant performance boost, with 44.0\% and 31.4\% enhancements over minRTT and BLEST, respectively. Remarkably, even under more challenging conditions, including a higher loss rate of 0.5\% and larger file sizes of 150MB, ALCS continues to demonstrate a commendable performance increase, providing a 24.8\% improvement over minRTT and a 9.8\% advantage over BLEST. This consistent superior performance across varying conditions underscores the robustness and efficiency of the ALCS algorithm.

\subsection{Retransmission rate of the schedulers}

\begin{figure}[t!]
    \centering
\includegraphics[width=0.7\linewidth]{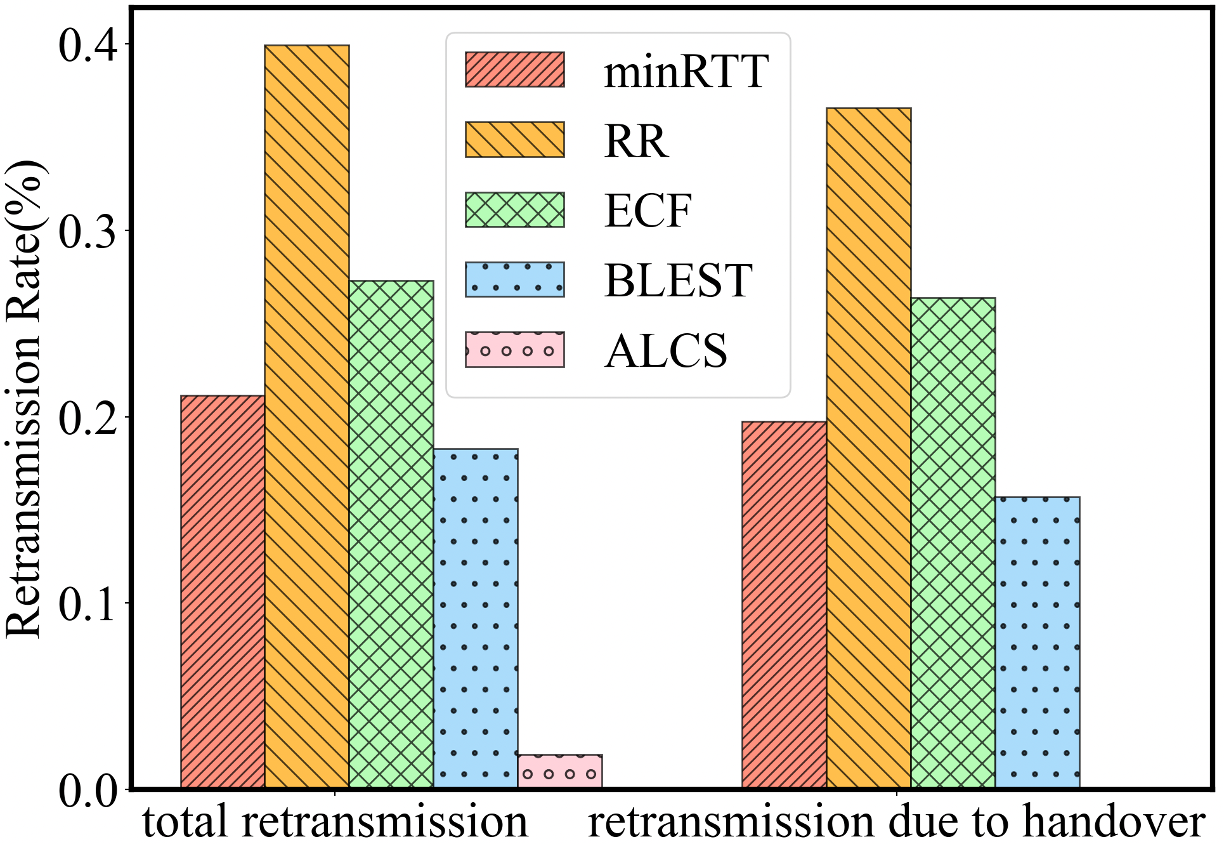}
    \caption{The comparison of retransmission rate with different schedulers.}
    \label{fig:retrans_rate}
    \vspace{-10pt}
\end{figure}

Furthermore, ALCS stands out by achieving the lowest overall retransmission rate and the fewest handover-induced retransmission rate compared to other schedulers, as illustrated in Fig. \ref{fig:retrans_rate}. Impressively, ALCS eliminates packet retransmissions caused by satellite handovers entirely. 

This impressive result is due to the proactive handover management mechanism integrated into ALCS. The algorithm preemptively deactivates the satellite path one RTT before a handover occurs and keeps it inactive until the handover process is complete. This strategy effectively minimizes the disruption caused by satellite handovers, reducing the need for retransmissions and enhancing the reliability of end-to-end communications.

In the STIN context, where retransmissions are predominantly triggered by satellite handovers, ALCS's approach addresses this issue directly, ensuring more stable and efficient data transmission.


\subsection{Fraction of data transferred over the satellite path}

\begin{figure}[t!]
\centering
\vspace{-12pt}
\subfloat[The loss rate of satellite network is set to 0.01\%.]{
    \includegraphics[width=0.7\linewidth]{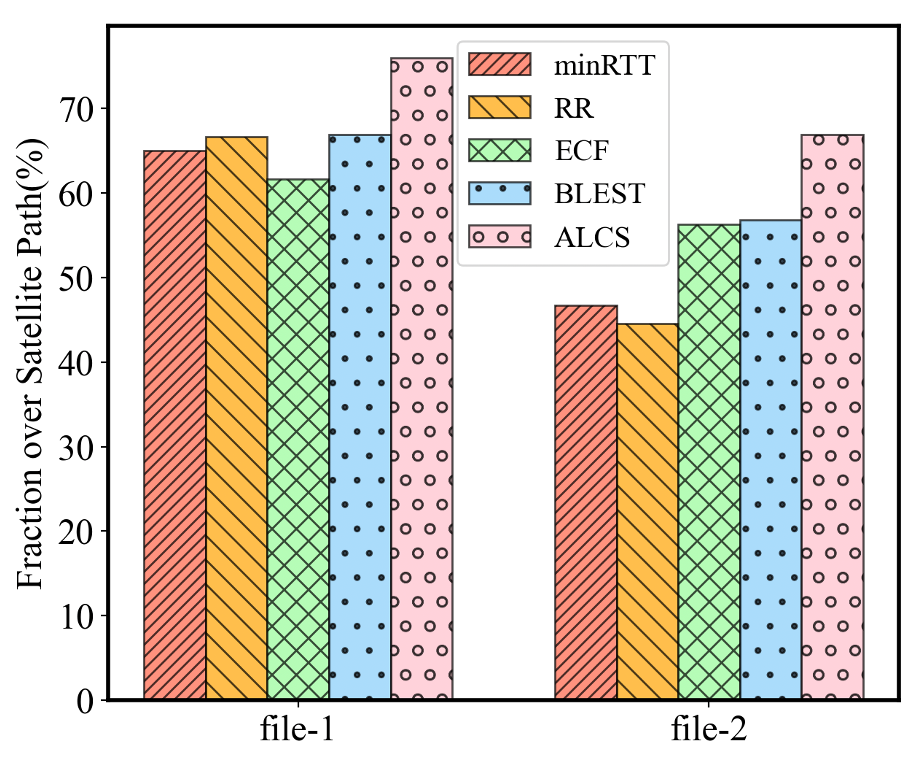}\label{fig:frac-over-sat-0.01}}\\
\subfloat[The loss rate of satellite network is set to 0.5\%.]{
    \includegraphics[width=0.7\linewidth]{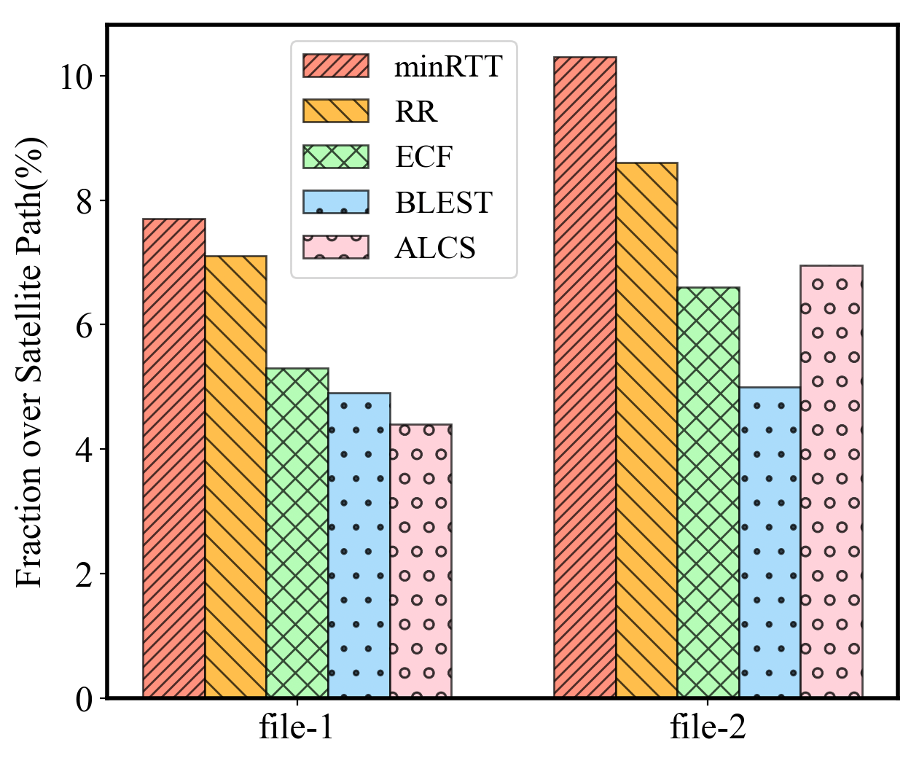}\label{fig:frac-over-sat-0.5}}
\caption{Fraction of Data Allocated to the Satellite Path based on Different Satellite Network's Loss Rates.}
\label{fig:frac-over-sat}
\vspace{-10pt}
\end{figure}

Fig. \ref{fig:frac-over-sat} displays the proportion of traffic allocated to the satellite path, which has a lower RTT, under varying satellite network loss rates of 0.01\% and 0.5\%. The results clearly demonstrate that the ALCS algorithm is more aggressive in exploiting the satellite path in comparison to other schedulers. This is particularly evident when the satellite link's condition is favorable. ALCS effectively leverages the satellite path's advantages to enhance data transmission efficiency when it outperforms the terrestrial path.

In contrast, in scenarios where the satellite path's loss rate significantly exceeds that of the terrestrial path, traditional schedulers like minRTT and RR  fail to adjust their allocation strategies effectively in response to the prevailing network conditions. On the other hand, ECF, BLEST, and ALCS adapt the proportion of traffic sent over the satellite path to account for the higher loss rate, thereby optimizing the overall performance by leveraging the strengths of each available network path. 

Notably, BLEST shows a tendency to favor the terrestrial path slightly more than the ECF scheduler in both the file-1 and file-2 transfer scenarios. Conversely, ALCS does not exhibit a fixed preference for either path. It demonstrates flexibility in its path selection, utilizing the satellite path the least during the file-1 transfer but not showing the same bias during the file-2 transfer. 
This adaptability indicates that ALCS's decision-making is influenced by the specific conditions and requirements of each transfer scenario, rather than adhering to a uniform strategy. This flexibility is a key feature of ALCS, allowing it to maximize overall efficiency under varying network conditions. 

\subsection{Performance analysis with multiple subflows}

\begin{figure}[t!]
\centering
\vspace{-10pt}
\subfloat[Transfer completion time.]{
    \includegraphics[width=0.7\linewidth]{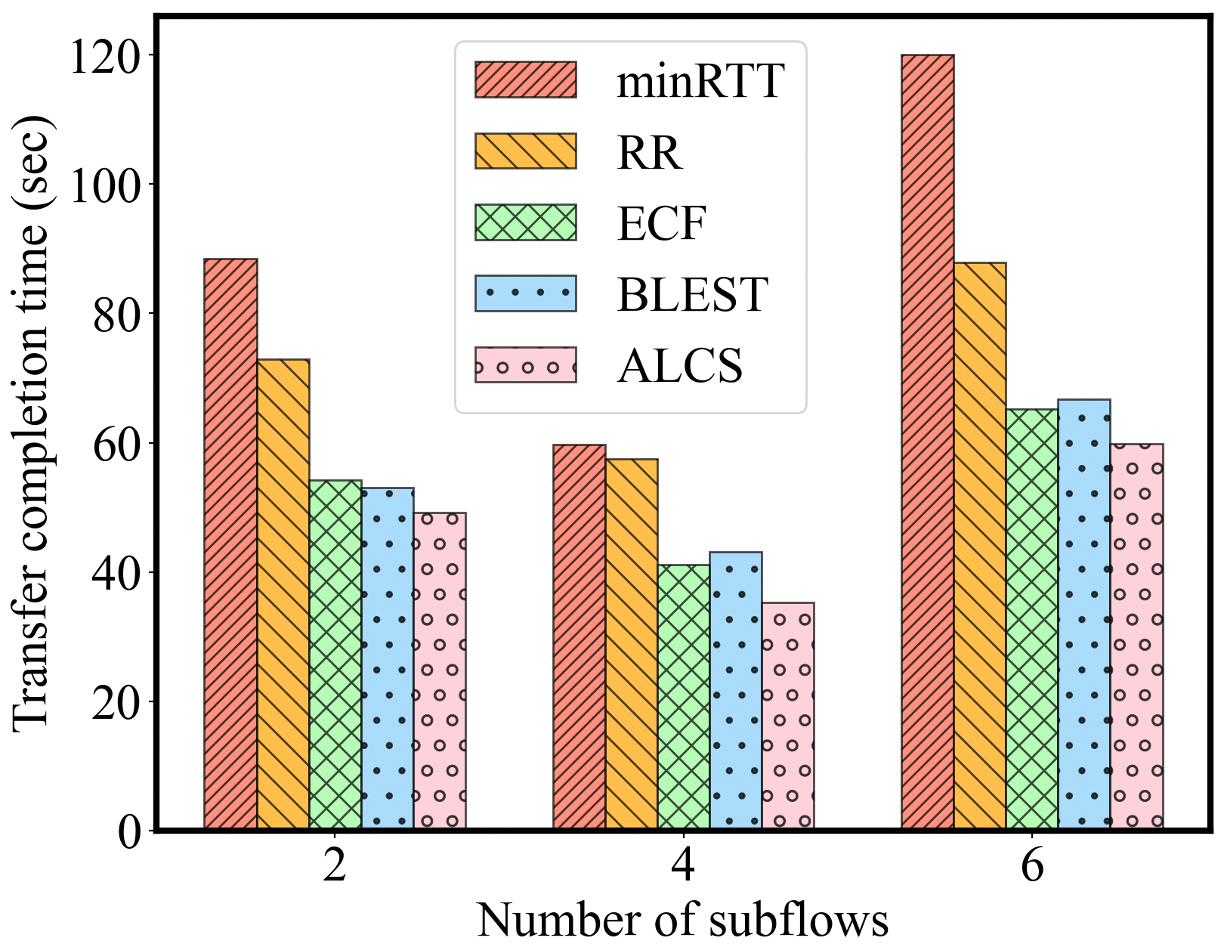}\label{fig:multi-sf-time}}\\
\subfloat[Average throughput.]{
    \includegraphics[width=0.7\linewidth]{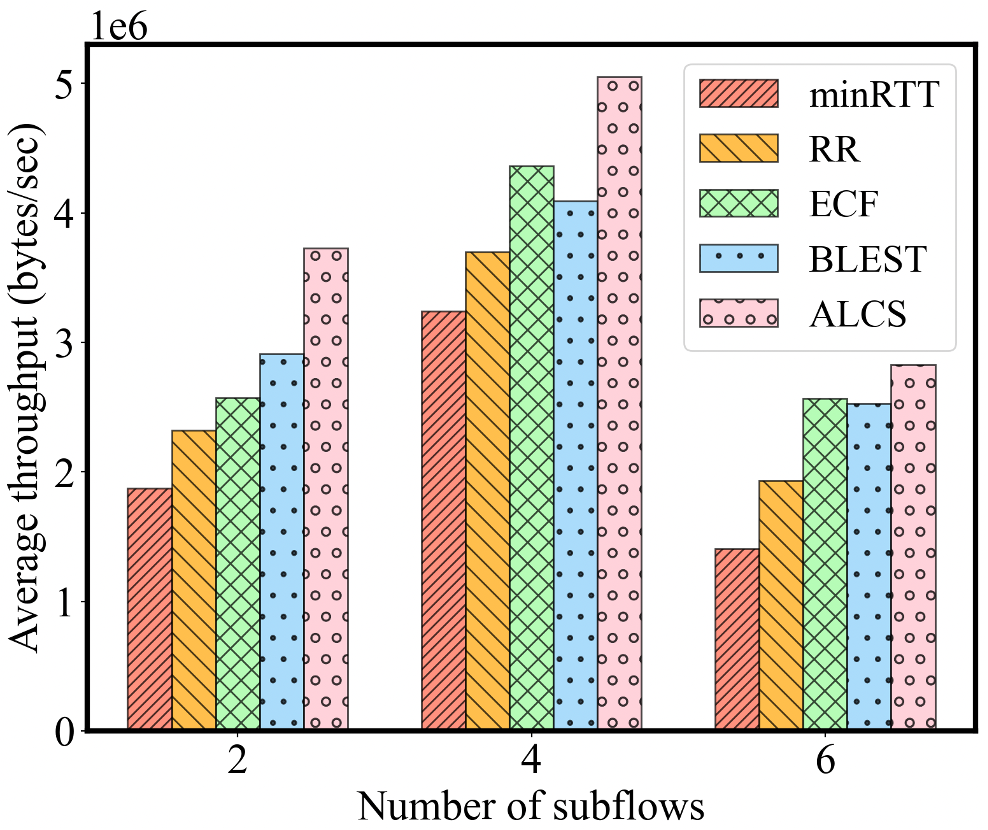}\label{fig:multi-sf-throughput}}\\
\caption{Performance comparison of various schedulers under different numbers of subflows.}
\label{fig:performance-multiple-subflows}
\vspace{-10pt}
\end{figure}

We proceed to examine the impact of varying the number of subflows, as depicted in Fig. \ref{fig:performance-multiple-subflows}. This analysis explores how the number of subflows influences the overall performance and behavior of the MPTCP schedulers. The performance evaluation is conducted during the transfer of a 150MB file within the STIN environment, where the satellite network loss rate is set to 0.5\% while the terrestrial network loss rate is fixed at 0.01\%.

Fig. \ref{fig:multi-sf-time} compares transfer completion time across different schedulers, and Fig. \ref{fig:multi-sf-throughput} highlights their performance in terms of average throughput. As the number of subflows increases from 2 (1 satellite and 1 terrestrial path) to 4 (2 satellite and 2 terrestrial paths), there is a notable improvement in the overall performance of all MPTCP schedulers. With more subflows, the schedulers have access to a broader selection of network paths, allowing more efficient distribution of data load and better aggregation of bandwidth across the available subflows. For instance, if one of the satellite paths experiences a connection disruption, the scheduler can seamlessly switch to an alternative satellite path, maintaining end-to-end connectivity and minimizing communication interruptions. Moreover, the presence of additional subflows facilitates more effective load balancing, allowing data to be allocated dynamically across paths based on real-time network conditions.

However, further increasing the number of subflows from 4 to 6 results in a degradation of system performance. This decline can be attributed to the increased complexity of scheduling, load balancing, and congestion control across multiple paths. With more available active subflowsk, optimizing their usage becomes more challenging, potentially leading to suboptimal utilization of network resources. Notably, the issue of out-of-order packet delivery worsens when 6 subflows are active, suggesting that beyond a certain point, adding more subflows introduces more complexity than benefit, ultimately reducing overall system efficiency.

Despite these challenges, the proposed ALCS consistently outperforms other schedulers in all tested scenarios, excelling in both transfer completion time and average throughput. This highlights ALCS’s superior adaptability and efficiency, even in the presence of increasing subflow complexity.

\section{Discussion and Future Work}\label{discussion}

\textbf{Applicability to diverse networks.}
ALCS demonstrates exceptional adaptability to various traffic patterns, making it effective not only within the STIN framework for file transmission and data segment scheduling but also across diverse network environments. Its utility extends to common terrestrial networks like Wi-Fi and LTE, enhancing performance under varying conditions. With its capability to efficiently handle different transfer sizes, ALCS significantly improves performance for numerous applications. For instance, it reduces latency and buffering in audio and video streaming, ensuring smoother web browsing experiences. Additionally, ALCS excels in managing large file transfers, making it valuable for software updates, cloud storage synchronization, and media distribution. This versatility positions ALCS as a robust solution for a wide range of network-based demands.


\textbf{Integration with MPTCP.} The core logic of ALCS has been implemented as a new scheduling module in the MPTCP Linux Kernel as mentioned in Section \ref{testbed}. This implementation ensures the seamless integration of ALCS within the MPTCP framework, leveraging the existing functionality and infrastructure of the MPTCP protocol. Moreover, the MPTCP framework provides a scalable and extensible foundation, allowing the ALCS scheduling module to be efficiently deployed and scaled to accommodate varying network loads and workloads. This seamless integration enhances the robustness and flexibility of ALCS in real-world applications. 

\textbf{Limitation and Future Work.} The current ALCS design is primarily focused on optimizing transfer completion time, a critical factor that directly impacts user experience, particularly for latency-sensitive applications. While reducing transfer delays can significantly enhance user satisfaction, it is acknowledged that other performance metrics, such as fairness and energy efficiency, may also be essential in certain contexts. As such, future iterations of the ALCS scheduler would aim to balance and optimize these additional metrics, adopting a more holistic approach to improving user experience. This would involve not only optimizing speed but also ensuring the equitable distribution of network resources and minimizing energy consumption.

Building on existing efforts, exploring the interplay between congestion control and packet scheduling across multiple paths in the STIN environment holds significant promise. This investigation is a natural extension of ALCS's current design, which already considers key path characteristics such as RTT, CWND, and in-flight packets—critical factors in developing effective congestion control algorithms.
While we currently use established algorithms like Cubic, along with multipath-specific methods such as Balia \cite{balia} and Olia \cite{olia}, our findings indicate that these approaches yield suboptimal performance in STIN. 
Therefore, it is imperative to develop a new congestion control algorithm specifically tailored to the unique challenges of STIN. Such an algorithm, integrated with ALCS, would further enhance multipath transmission performance and resilience, ensuring robust, efficient data transfer even in highly dynamic network conditions.

\section{Related Work} \label{related-work}

\textbf{Multipath extensions over TCP.}
MPTCP is a significant extension of traditional TCP, designed to leverage multiple network paths between two end hosts for transferring a single data stream. Standardized by the Internet Engineering Task Force in 2013~\cite{rfc8684}, MPTCP has been integrated into various operating systems, including Linux~\cite{mptcp-linux-kernel}, FreeBSD~\cite{freebsd-mptcp}, and Solaris~\cite{solaris-mptcp}. Over the years, it has become a prominent research area, with ongoing efforts from academia and industry to enhance its performance and functionality. Key focus areas include optimizing path management, improving data scheduling, and refining congestion control mechanisms.




\textbf{Packet scheduling in MPTCP.} 
Packet scheduling is a crucial element that significantly impacts the performance of MPTCP transport systems. In recent years, substantial research has been devoted to advancing packet scheduling algorithms for MPTCP. Kuhn et al. \cite{daps} propose a delay-aware packet scheduling algorithm to handle path heterogeneity in forward delay and CWND, ensuring in-order packet delivery. Similarly, Yang et al.~\cite{otia} introduce a scheduler that predicts the arrival time of each packet at the receiver and prioritizes subflows based on the shortest predicted arrival time. Other schedulers such as STMS~\cite{stms} and DEMS~\cite{dems} have also been developed to reduce out-of-order arrivals. STMS prioritizes the faster subflow for packet allocation and uses the slower subflow to transmit packets with larger sequence numbers. DEMS splits a data chunk across two subflows and strategically decouples the paths for chunk delivery efficiency.
Similar to ECF, STTF, proposed by Hurtig et al. \cite{sttf}, aims to reduce the transmission time of each data segment by considering the current congestion state of each subflow, and dynamically adjusting to network conditions. Moreover, several loss-aware schedulers integrate loss rates into their decision-making process for selecting subflows and prioritizing paths with the lowest estimated delay for segment transmission~\cite{lamps, late}. Various machine learning-based approaches have also been applied to MPTCP packet scheduling~\cite{olaps, reles, peekaboo}. 

\textbf{Multipath transmission within STIN.} The application of multipath transmission protocols such as MPTCP and multipath QUIC \cite{de2017multipath} (MPQUIC) in the context of STIN remains a relatively underexplored area compared to their extensive application in terrestrial networks. Xiao et al.~\cite{xiao2022efficient} introduce an MPTCP scheme based on an SDN architecture to enhance the throughput and efficiency in STIN, while Wang et al.~\cite{cdmr} propose a computing-dependent multipath routing paradigm to optimize task computing during end-to-end transmission. Yang et al. develop a mobility-aware MPQUIC framework, along with customized data scheduling~\cite{mams-quic} and congestion control algorithms~\cite{ma-quic}, specifically tailored to the unique end-to-end path conditions within STIN.
The limited research on multipath transmission in STIN highlights the need for further exploration and innovative solutions. 

\section{Conclusion} \label{conclusion}

In this work, we present an analytical framework for MPTCP-enabled STIN and demonstrate the performance degradation of several traditional existing MPTCP schedulers when deployed in STIN environments. To address the three key challenges posed by STIN—severe path heterogeneity, high volatility, and frequent handovers—we introduce a STIN-optimized multipath scheduling approach called ALCS. ALCS addresses path heterogeneity by integrating RTT, CWND, inflight packets, and queuing delay to estimate transmission delays across subflows. These estimates are further refined using latency compensation mechanisms to mitigate the unpredictable nature of satellite links. In addition, ALCS incorporates a proactive handover management strategy that leverages the predictability of satellite movement to effectively manage frequent handovers.
Through experiments conducted in our custom-built testbed, we evaluate ALCS against the minRTT, RR, ECF, and BLEST schedulers, focusing on metrics such as transfer completion time, average throughput, and retransmission rate under various network conditions. The results consistently demonstrate the superior performance of ALCS, delivering improvements ranging from 9.8\% to 44.0\% over the baseline algorithms across different scenarios.

\bibliographystyle{IEEEtran}
\bibliography{reference}

\end{document}